\documentclass[10pt]{iopart}
\usepackage[utf8]{inputenc}
\usepackage[english]{babel}
\usepackage{float}
\expandafter\let\csname equation*\endcsname\relax
\expandafter\let\csname endequation*\endcsname\relax
\usepackage{amsmath}
\usepackage{graphicx}
\usepackage{subcaption}
\usepackage[normalem]{ulem}
\usepackage{xcolor}
\begin{document}

\title[Positive streamers in air at different electric fields]{Simulations of positive streamers in air in different electric fields: steady motion of solitary streamer heads and the stability field}

\author{Hani Francisco$^1$, Jannis Teunissen$^1$, Behnaz Bagheri$^2$, Ute Ebert$^{1,2}$}
\address{$^1$ Multiscale Dynamics Group, Centrum Wiskunde \& Informatica (CWI), Amsterdam, The Netherlands}
\address{$^2$ Department of Applied Physics, Eindhoven University of Technology, Eindhoven, The Netherlands}

\begin{abstract}
  We simulate and characterize positive streamers in ambient air in homogeneous background electric fields from $4.5$ to $26$~kV/cm in a $4$~cm gap. They can accelerate or decelerate depending on the background electric field. Many experiments have shown that a streamer keeps propagating in a stable manner in the so-called stability field of $4.5$ to $5$~kV/cm. Our fluid streamer simulations in STP air show that: (1) In a homogeneous field larger than $4.675$~kV/cm, a single streamer accelerates, and in a lower field, it decelerates and eventually stagnates with a small radius and very high field enhancement. (2) In a field of $4.675$~kV/cm, the streamer head propagates with an approximately constant velocity of $6.7 \times 10^4$~m/s and an optical radius of $55~\mu$m over distances of several centimeters as a stable coherent structure. These values for the radius and velocity agree well with measurements of so-called minimal streamers. (3) Behind the uniformly translating streamer head, the channel conductivity decreases due to electron attachment and recombination, and the electric field returns to its background value about $1$~cm behind the head. The propagation behavior of the solitary streamer agrees with the original definition of the stability field, which is the homogeneous field in which a streamer can propagate with a constant speed and shape.
\end{abstract}

Revised manuscript resubmitted to \PSST on Sept.~30, 2021.

\maketitle

\ioptwocol

\section{Introduction}

Streamer discharges are transient discharges that serve as precursors to other gas discharges such as sparks and lightning leaders. They are rapidly growing ionized channels that are characterized by a curved space charge layer around their plasma body, which screens the electric field in their interior and enhances it ahead of them~\cite{Babaeva1997, Kulikovsky1998, Pancheshnyi_2005, ebert_2010, Nijdam_review_2020}. The enhanced field in the active zone at the streamer head exceeds the electric breakdown value, and the multiplication of electrons in this region drives the propagation of the streamer. Streamers have multiple applications in various fields, including, but not limited to, medicine~\cite{laroussi2014}, combustion~\cite{starikovskaia_plasma-assisted_2014}, and surface treatments~\cite{bardos2010}.

Streamers can form even if the background electric field is below breakdown as long as there is an area where the field is enhanced above the breakdown threshold. This allows for the observation of streamers in a wide range of electric fields in the laboratory~\cite{Allen_1995, phelps_fieldenchanced, briels_positive_2008}. Numerically, it has been a challenge to study streamers in low background electric fields due to issues related to streamer initiation and streamer branching~\cite{Luque_branching_2011}. In \cite{Pancheshnyi_2004}, a streamer simulation was performed with a low background electric field, and that led to the first study of streamer stagnation dynamics. This was studied more recently in \cite{starikovskiy2021} where decelerating streamers were obtained by having inhomogenous gas density.

In this paper, we employ an approach that allows us to look at streamers propagating in low background fields: we initiate the streamer in a higher field, let it propagate for some time, and then reduce the background electric field to a much lower value. This scheme allows us to do a parameter sweep of background electric fields farther below electric breakdown, going as low as $4.5$~kV/cm.

In a recent paper~\cite{francisco2021}, we have studied single positive streamers in dry air in a homogeneous background electric field of $15$~kV/cm, about half the breakdown field, at standard temperature and pressure. The radius and the velocity of the streamers increased with the streamer length, as observed by many authors before. When the electron attachment rate was artificially increased in regions below electric breakdown, we found that with increasing attachment rate, streamer velocities and radii could grow less, not at all, or even decrease. Additionally, streamer heads could keep propagating even if the conductivity of the streamer channels was already negligible a short distance behind the streamer head. We did not specify gases where such dynamics could actually be observed.

In the current work, we show that the same variation of streamer dynamics can occur in ambient air by simply decreasing the homogeneous background electric field. We find that for a background field of about $4.675$~kV/cm, the streamer head propagates with a constant radius and velocity. The current that flows through the streamer channel is already negligible close behind the head - the electric field returns to the background field value at the back of an electrically isolated streamer head. If the background electric field is even smaller, the streamer velocity and radius decrease while the maximal electric field at the head rapidly increases, and this could go on until the streamer stops. Finding uniform streamer propagation in STP air confirms the old concept of the stability field~\cite{phelps_fieldenchanced, griffiths_effectofairpressure, gallimberti_longspark} that is frequently used in high voltage engineering but had little support up to now from fundamental physical modeling.

The paper is structured as follows. Details about the numerical modeling are presented in section~\ref{sec:model}, where the computational domain is described along with the initial conditions of the simulations in section~\ref{sec:init}. Section~\ref{sec:results} features and discusses the results of our simulations. In section~\ref{sec:2cases}, we present the case of a uniformly translating streamer in ambient air together with the more familiar case of an accelerating streamer, and in section~\ref{sec:behaviors}, we show how streamer behaviour more generally depends on the background electric field. We also include decelerating streamers in that section. Section~\ref{sec:validation} has comparisons between our simulation results and experimental measurements, and we discuss there the original concept of the stability field and its connection to our solitary streamers. We conclude in section~\ref{sec:conclusion}, where we summarize our results and communicate ideas for future studies.

\section{Discharge Model\label{sec:model}}

\subsection{Model equations and reactions}

\begin{table}
    \centering
    \begin{tabular}{c c c}
        \hline
        1 & $e + {\rm N}_2 \xrightarrow{} 2 e + {\rm N}_2^+$ & $k_1 \left( E/N \right)$\\
        2 & $e + {\rm O}_2 \xrightarrow{} 2 e + {\rm O}_2^+$ & $k_2 \left( E/N \right)$\\
        3 & $e + {\rm O}_2 + {\rm O}_2 \xrightarrow{} {\rm O}_2^- + {\rm O}_2$ & $k_3 \left( E/N \right)$\\
        4 & $e + {\rm O}_2 \xrightarrow{} {\rm O} + {\rm O}^-$ & $k_4 \left( E/N \right)$\\
        5 & ${\rm M} + {\rm O}_2^- \xrightarrow{} e + {\rm O}_2 + {\rm M}$ & $k_5 \left( E/N \right )$\\
        6 & ${\rm N}_2 + {\rm O}^- \xrightarrow{} e + {\rm N}_2{\rm O}$ & $k_6 \left( E/N \right)$\\
        7 & ${\rm O}_2 + {\rm O}^- \xrightarrow{} {\rm O}_2^- + {\rm O}$ & $k_7 \left ( E/N \right)$\\
        8 & ${\rm O}_2 + {\rm O}^- + {\rm M} \xrightarrow{} {\rm O}_3^- + {\rm M}$ & $k_8 \left ( E/N \right)$\\
        9 & ${\rm N}_2^+ + {\rm N}_2 + {\rm M} \xrightarrow{} {\rm N}_4^+ + {\rm M}$ & $k_9$\\
        10 & ${\rm N}_4^+ + {\rm O}_2 \xrightarrow{} 2 {\rm N}_2 + {\rm O}_2^+$ & $k_{10}$\\
        11 & ${\rm O}_2^+ + {\rm O}_2 + {\rm M} \xrightarrow{} {\rm O}_4^+ + {\rm M}$ & $k_{11}$ \\
        12 & $e + {\rm O}_4^+ \xrightarrow{} 2 {\rm O}_2$ & $k_{12} \left ( E/N \right)$ \\
        \hline
    \end{tabular}
    \caption{List of reactions included in the model. M stands for both ${\rm O}_2$ and ${\rm N}_2$, and $E/N$ is the reduced electric field calculated from the electric field $E$ and the gas density $N$. The electron impact reactions $1-4$ have reaction rate coefficients calculated with Bolsig+~\cite{hagelaar_solving_2005} while the reaction rate coefficients of the ion reactions $5-11$ were taken from \cite{Pancheshnyi_effective_2013, Aleksandrov_ionization_1999}. The reaction rate coefficient of reaction 12 is calculated~\cite{kossyi_kinetic_1992} from the mean electron energy calculation of Bolsig+.}
    \label{tab:rxns}
\end{table}

We used a plasma fluid model with local field approximation to simulate positive streamers in artificial dry air at standard temperature and pressure at different homogeneous background electric fields. The model equations, transport coefficients, and included reactions and reaction rate coefficients are the same as in our earlier paper~\cite{francisco2021}. 

The electron density $n_e$ evolves in time according to the equation
\begin{equation}
    \frac{\partial{n_e}}{\partial{t}} = \nabla \cdot \left( n_e \mu_e \textbf{E} + D_e \nabla n_e  \right) + S_i - S_\eta + S_{ph} + S_{ion},
\end{equation}
where $\mu_e$ is the electron mobility, $\textbf{E}$ is the electric field, $D_e$ is the electron diffusion coefficient, $S_i$ is the impact ionization source term, $S_\eta$ is the electron attachment source term, $S_{ph}$ is the non-local photoionization source term, and $S_{ion}$ is the source term for electron detachment reactions minus the electron-ion recombination reaction. Table~\ref{tab:rxns} summarizes the reactions incorporated in the model.

We used the reactions given in \cite{luque2017streamer} excluding the ion-ion recombination reactions and the reactions that involved water. This chemical model is based on \cite{Pancheshnyi_effective_2013, Aleksandrov_ionization_1999, kossyi_kinetic_1992} and focuses on the electron density evolution, in accordance with our focus on the conductivity inside the streamer channel.

Nearly all reaction rate coefficients in Table~\ref{tab:rxns} are a function of the reduced electric field, and only reactions~$9$-$11$ have constant reaction rate coefficients. The electron Boltzmann equation solver Bolsig+~\cite{hagelaar_solving_2005} was utilized under the assumption of spatially dependent electron density evolution to calculate the reaction rate coefficients for the electron impact reactions and the transport coefficients $\mu_e$ and $D_e$ using electron-neutral scattering cross sections obtained from the Phelps database~\cite{phelps_anisotropic_1985,phelps_data} retrieved in March 2019.

The source terms for impact ionization, electron attachment, and electron detachment minus electron-ion recombination are computed using
\begin{equation}
    S_i = k_1 n_e \left[ {\rm N}_2 \right] + k_2 n_e \left[ {\rm O}_2 \right], 
\end{equation}
\begin{equation}
    S_\eta = k_3 n_e \left [ {\rm O}_2 \right]^2 + k_4 n_e \left[ {\rm O}_2 \right].
\end{equation}
\begin{equation}
    S_{ion} = k_5 \left [ {\rm M} \right]\left [ {\rm O}_2^{-} \right] + k_6  \left[ {\rm N}_2 \right]\left[ {\rm O}^{-} \right] - k_{12}  n_e \left[ {\rm O}_4^{+} \right],
\end{equation}
where [Z$_i$] stands for the density of the species Z$_i$, and [M] = [N$_2$] + [O$_2$]. [N$_2$] and [O$_2$] are assumed to be constant in our simulations as the degree of ionization within streamers at standard temperature and pressure is small.

The photoionization source term is given by
\begin{equation}
    S_{ph}({ \bf r})=\int d^3 r'\;\frac{I( {\bf r'}) f(|{\bf r}-{\bf r'}|)}{4 \pi |{\bf r} - {\bf r'}|^2}
\label{equ:photo}
\end{equation}
where $I\left({\bf r}\right)$ is the source of ionizing photons, $f(r)$ is the absorption function, and $4\pi |{\bf r} - {\bf r'}|^2$ is a geometric factor. Following Zheleznyak's model~\cite{zheleznyak_photoionization_1982}, the photon source term $I\left({\bf r}\right)$ is calculated using
\begin{equation} \label{eq:photoionization}
I\left(\textbf{r}\right) = \frac{ p_q}{p + p_q}\xi S_i\left(\textbf{r}\right)
\end{equation}
where $p$ is the actual gas pressure, $p_q$ is a gas-specific quenching pressure, and $\xi$ is a proportionality factor. In principle, this proportionality factor is field-dependent~\cite{zheleznyak_photoionization_1982}, but in this paper, we set it to $\xi = 0.075$. Furthermore, we use a quenching
pressure of $p_q = 40 \, \mathrm{mbar}$.
In Zheleznyak's model, $f(r)$ is an effective function for the absorption of photons in the wave length range of
$98$ to $102.5$~nm. It is obtained with
 \begin{equation}
f(r)=\frac{\exp(-\chi_{\mathrm{min}}p_{O_2}r)-\exp({-\chi_{\mathrm{max}}}p_{O_2}r)}{r\ln(\chi_{\mathrm{max}}/\chi_{\mathrm{min}})},
\label{equ:absorption-function}
\end{equation}
where $\chi_{\mathrm{max}}\approx1.5\times10^2/(\textnormal{mm bar})$,
$\chi_{\mathrm{min}}\approx2.6/(\textnormal{mm bar})$, and $p_{O_2}$ is the
partial pressure of oxygen.
We used a set of Helmholtz differential equations~\cite{bourdon_efficient_2007,luque_photoionization_2007} with Bourdon's three-term parameters~\cite{bourdon_efficient_2007} to evaluate the photoionization integral.

The charged species ${\rm N}_2^+$, ${\rm N}_4^+$, ${\rm O}_2^+$, ${\rm O}_4^+$, ${\rm O}^-$, ${\rm O}_2^-$, and ${\rm O}_3^-$, and the neutral species O and ${\rm N}_2{\rm O}$ evolve in time according to the continuity equation
\begin{equation}
\frac{\partial \left[ Z_i \right ]}{\partial t} =  - s_i \nabla \cdot \left( \left[ Z_i \right ] \mu_i \textbf{E} \right) + S_{Z_i}
\label{eq:ion_cont}
\end{equation}
where $s_i = \pm 1$ is the sign of the electric charge of species $i$ and $\mu_i$ is their mobility. Since ion mobilities are typically about two orders of magnitude lower than electron mobilities, we neglect ion motion for simplicity in most of this paper. However, we investigate the effect of ion motion in section~\ref{sec:ion_motion}, in which all ion mobilities are set to 2.2 $\times 10^{-4}$ m$^2/$V s \cite{Tochikubo_2002}. Finally, neutral species are always immobile in our simulations.

Calculations for the electric potential $\phi$ and the electric field use the equations
\begin{equation}
\nabla^2 \phi = - \frac{\rho}{\epsilon_0},\quad \textbf{E} = - \nabla \phi,
\end{equation}
where $\rho$ is the space charge density and $\epsilon_0$ is the vacuum permittivity. The space charge density is calculated using $\rho = e \left ( n_i - n_e \right)$ where $e$ is the elementary charge and $n_i$ is the density of all positive ions minus the density of all negative ions. 

\subsection{Computational method and domain \label{sec:domain}}

The simulations were run using Afivo-streamer~\cite{teunissen_simulating_2017, Teunissen_afivo_2018}, a simulation tool for plasma fluid models that uses geometric multigrid techniques, an octree-based adaptive mesh refinement system, and OpenMP parallelization. The present results are for single streamers, and these assume that they are cylindrically symmetric. This allows the calculation to be performed effectively in just the two coordinates $r$ and $z$.

Our computational domain in this study is cylindrically symmetric and has a length of $40$~mm and a radius of $20$~mm. To disregard boundary effects, the simulation is set to end once the streamer head is within $10$~mm from the opposite end of the domain. The streamer head position is identified as the point where the electric field is maximum in the domain.

The electric potential was fixed at $z = 0$~mm and $z = 40$~mm to achieve a homogeneous background electric field pointing in the $-\hat{z}$ direction. At $r =20$~mm, Neumann zero boundary conditions ($\partial_r \phi = 0$) were applied on the electric potential, and for $r = 0$~mm, the boundary condition follows from cylindrical symmetry. Neumann zero boundary conditions are applied for the electron density at all boundaries, and no background ionization was introduced into the domain.

We used the same refinement criteria as described in ~\cite{francisco2021}: 
Adaptive mesh refinement is employed with the grid set to have a minimum size of $2.4~\mu$m.
The refinement and derefinement criteria are based on the local electric field value as in \cite{teunissen_simulating_2017} with an additional criterion based on the charge density: refine if $\alpha(1.2\times E) \Delta x > 0.5 $ and derefine if both $\alpha(1.2\times E)\Delta x < 7.5 \times 10^{-2}$ and $|\rho|/\epsilon_0 < 9.0 \times 10^{10}~\rm{V/m^2}$, where $\alpha(E)$ is the field-dependent ionization coefficient, $E$ is the electric field strength, and $\Delta x$ is the grid spacing.
To obtain a clearer picture of the equipotential lines in the regions behind the streamer head, we modified our derefinement criterion for the streamers with a background field of $4.65$~kV/cm and below so that derefinement stops when the cell width gets to $4~\mu$m.

\subsection{Initial conditions \label{sec:init}}

\begin{figure}
    \centering
    \includegraphics[width=\linewidth]{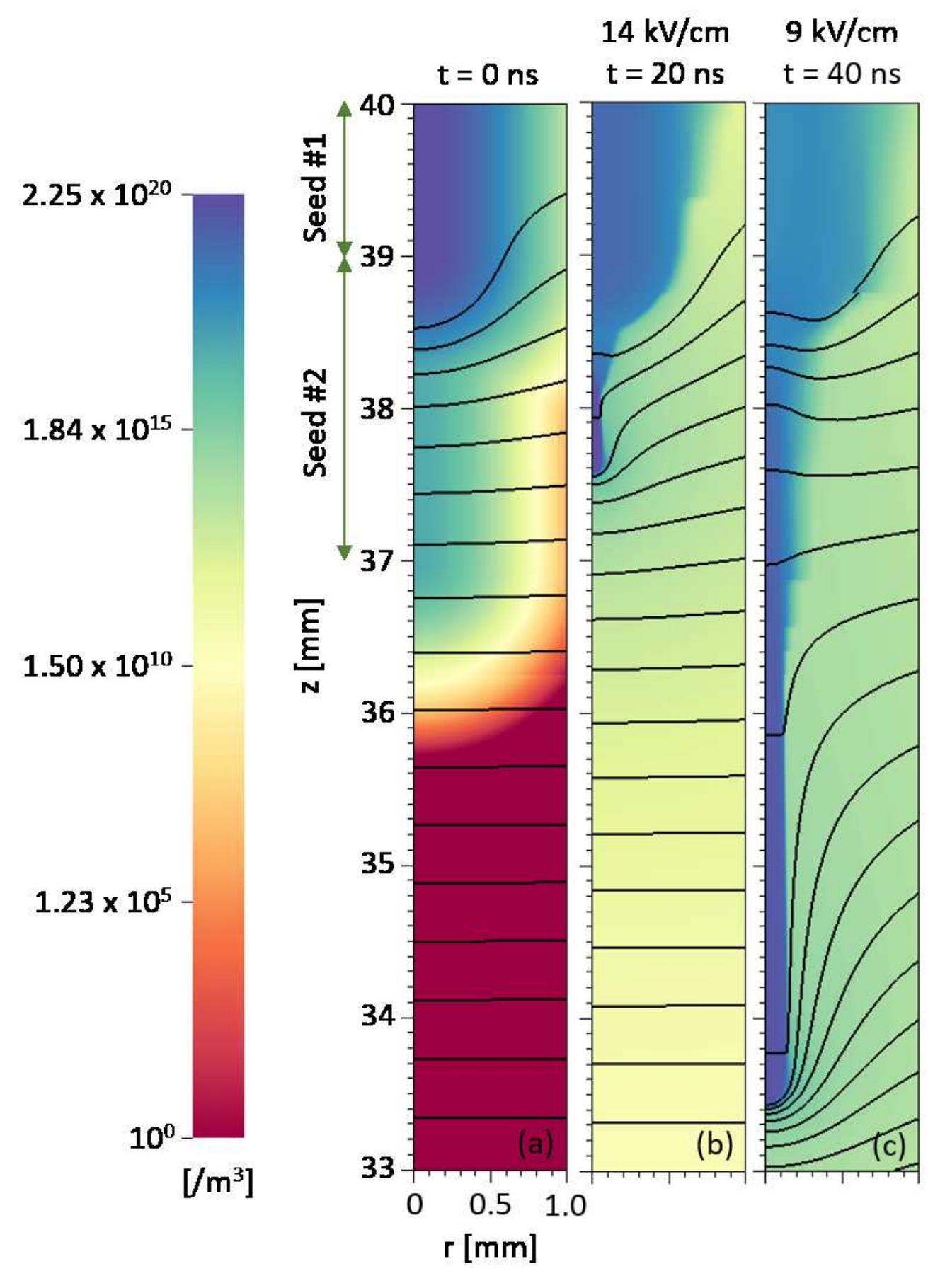}
    \caption{Initial conditions for the different background electric fields used in this paper. Shown are contour plots for the electron number densities together with black equipotential lines. The leftmost panel (a) is for streamers in background fields of $14$~kV/cm and higher, the middle panel (b) is for streamers with background fields below $14$~kV/cm to $9$~kV/cm, and the rightmost panel (c) is for streamers in fields below $9$~kV/cm. Note that the computational domain extends from $0$ to $40$~mm in the $z$ direction and $0$ to $20$~mm in the $r$ direction, and only a part of the domain is shown in this figure.}
    \label{fig:initial_conditions}
\end{figure}

{\bf For homogeneous background electric fields of at least $14$~kV/cm}, streamers easily initiate and propagate from a neutral seed of equal electron and positive ion densities, which we placed on the upper boundary of the domain, along the axis of symmetry. Another neutral seed is placed below the first seed to provide an initial source of electrons.
The first seed is $0.25$~mm wide, $1$~mm long, and has $2.25 \times 10^{20} \rm{/m}^3$ electrons and positive ions while the second seed is $0.2$~mm wide, $2$~mm long, and has $10^{17} \rm{/m}^3$ electrons and positive ions. Both seeds decay with a Gaussian profile. This set-up is illustrated in the left-most panel of figure~\ref{fig:initial_conditions}.

{\bf Single streamers are more difficult to obtain in lower background electric fields} because either the field enhancement proves to be insufficient for streamers to initiate or the streamer branches after propagating a short length. Branching breaks the cylindrical symmetry of a single streamer channel, and thus cylindrically symmetric simulations are not appropriate to describe such phenomena~\cite{luque_density_2012}. To investigate low-field streamers, a streamer is first initiated and allowed to propagate for some time in a higher background field before the background electric field is instantaneously reduced to a lower value. This approach allows us to study single continuously propagating and non-branching streamers in fields lower than $14$~kV/cm.

{\bf For electric fields from $9$~kV/cm to $12$~kV/cm}, a streamer was first initiated in a field of $14$~kV/cm and allowed to grow for $20$~ns before instantaneously reducing the background electric field. Thus, the low-field streamers grow from a streamer with a $53.5~\mu$m radius and head at $z = 37.6$~mm as shown in the middle panel of figure~\ref{fig:initial_conditions}.

{\bf For even lower fields}, this approach still encounters the same initiation and branching problems that were previously stated.
Thus, for streamers in background electric fields below $9$~kV/cm, we used the $9$~kV/cm streamer after $40$~ns of propagation as the initial condition, i.e. the field was reduced twice. First, the field was changed from $14$ to $9$~kV/cm after $20$~ns, and then it was modified further to the final electric field after $40$~ns. This gives a $155~\mu$m wide streamer with its head at $z = 33.4$~mm as the starting state for these lower field simulations. This initial condition can be seen in the right-most panel of figure~\ref{fig:initial_conditions}, which matches the left-most panel of figure~\ref{fig:electron-density}. This last approach allowed us to simulate single streamers in background electric fields as low as $4.5$~kV/cm.

\subsection{Calculation of optical radii}

\begin{table}
    \centering
    \begin{tabular}{c c c}
        \hline
        $1$  & ${\rm e} + {\rm N}_2 \xrightarrow{} {\rm e} + {\rm N}_2\left( {\rm C} ^3\Pi_u\right)$ & $k_{ex}\left( E/N \right)$  \\
        $2$  & ${\rm N}_2\left( {\rm C} ^3\Pi_u\right) + {\rm N}_2 \xrightarrow{} 2{\rm N}_2$  & $k_q^{{\rm N}_2}$ \\
        $3$  & ${\rm N}_2\left( {\rm C} ^3\Pi_u\right) + {\rm O}_2 \xrightarrow{} {\rm N}_2 + {\rm O}_2$  & $k_q^{{\rm O}_2}$ \\
        $4$  & ${\rm N}_2\left( {\rm C} ^3\Pi_u\right) \xrightarrow{} {\rm N}_2\left( {\rm B} ^3\Pi_g\right) + h\nu$  & $1/\tau_0$ \\
        \hline
    \end{tabular}
    \caption{Reactions to calculate the optical emission of streamers. Bolsig+~\cite{hagelaar_solving_2005} with the Phelps database~\cite{phelps_anisotropic_1985, phelps_data} was used to calculate for $k_{ex}\left( E/N \right)$, while $k_q^{{\rm N}_2} = 0.13 \times 10^{-10}$ cm$^3/{\rm s}$, $k_q^{{\rm O}_2} = 3.0 \times 10^{-10}$ cm$^3/{\rm s}$, and $\tau_0 = 42$~ns are from \cite{Pancheshnyi_2000}. Reaction~$4$ leads to the emission of optical photons with wavelength $337.1$~nm~\cite{Pancheshnyi_2000} or energy $3.7$~eV.}
    \label{tab:optic_rxns}
\end{table}

All radii given in the present paper are optical radii, as they would be measured experimentally. More precisely, this optical radius is half of the full width at half maximum (FWHM) of the calculated optical emission, in contrast to the definition of the streamer radius as the location of the maximum of the radial component of the electric field in previous papers~\cite{francisco2021, bagheriSimulationPositiveStreamers2020}. Four additional reactions were added to our model to incorporate the density of ${\rm N}_2\left( {\rm C} ^3\Pi_u\right)$, the excited state of $N_2$ responsible for most radiation in the visible spectral region~\cite{Pancheshnyi_2005}. These reactions are listed in Table~\ref{tab:optic_rxns} with their corresponding reaction rate coefficients.

We compute the optical radius from the density of ${\rm N}_2\left( {\rm C} ^3\Pi_u\right)$. A forward Abel transform was done on $\left[{\rm N}_2\left( {\rm C} ^3\Pi_u\right)\right]$ in cylindrical coordinates to get its 2D projection in Cartesian coordinates. From the 2D projection we only considered the area below $z = 33$~mm to disregard the effects of the seeds used for initiation. The densities were normalized and summed along the vertical axis, producing a 1D profile along the horizontal axis from where we searched for the maximum density. From the point of maximum density, the farthest coordinates in the horizontal direction where the density was at least half of the maximum density were identified, and the distance between these two identified points was regarded as the head diameter. Half of that value is the optical radius we report.

\section{Simulation results \label{sec:results}}

First, in section~\ref{sec:2cases}, we will discuss the particular cases of single streamers in a background field of 4.65 kV/cm and 14~kV/cm which are examples of solitary and accelerating streamers. Then we will look at streamer behavior as a function of the background field in section~\ref{sec:behaviors}.

\subsection{Solitary streamers and accelerating streamers \label{sec:2cases}}

\begin{figure*}
    \centering
    \includegraphics[width=\linewidth]{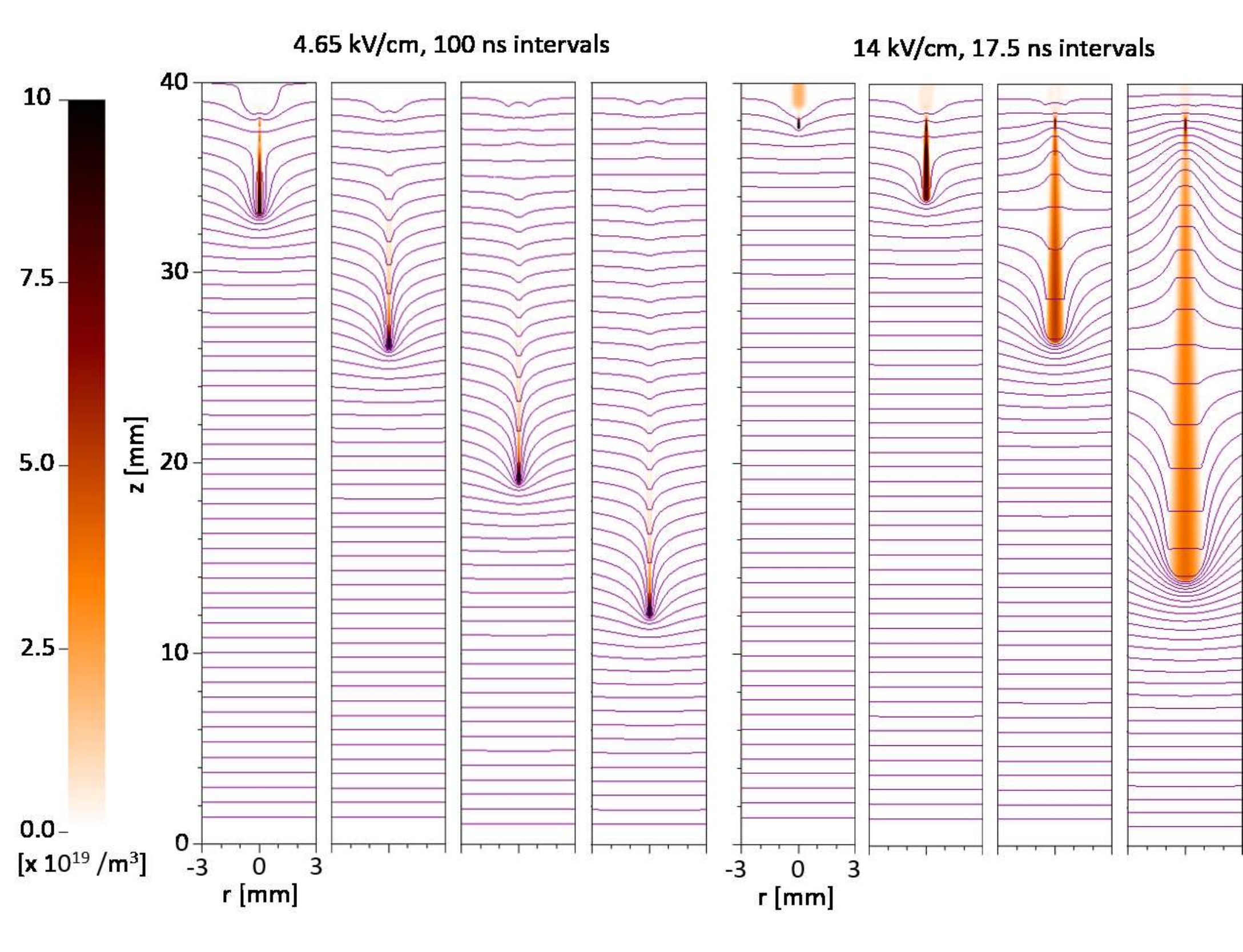}
    \caption{Time evolution of the electron density of streamers in air at different background electric fields. Purple equipotential lines are included. The panels of the $4.65$~kV/cm streamer differ by time steps of $100$~ns, while the $14$~kV/cm streamer is shown in time steps of $17.5$~ns. The full $z$ axis is shown, but the figure zooms into the radial region $r\le 3$~mm, while the full simulation domain extends up to $r = 20$~mm. Note that despite the limit in the color legend, the maximum electron density for the presented cases of the $4.65$~kV/cm streamer goes above $10 \times 10^{19}$ /m$^3$.}
    \label{fig:electron-density}
\end{figure*}

\begin{figure}
    \centering
    \includegraphics[width=\linewidth, keepaspectratio]{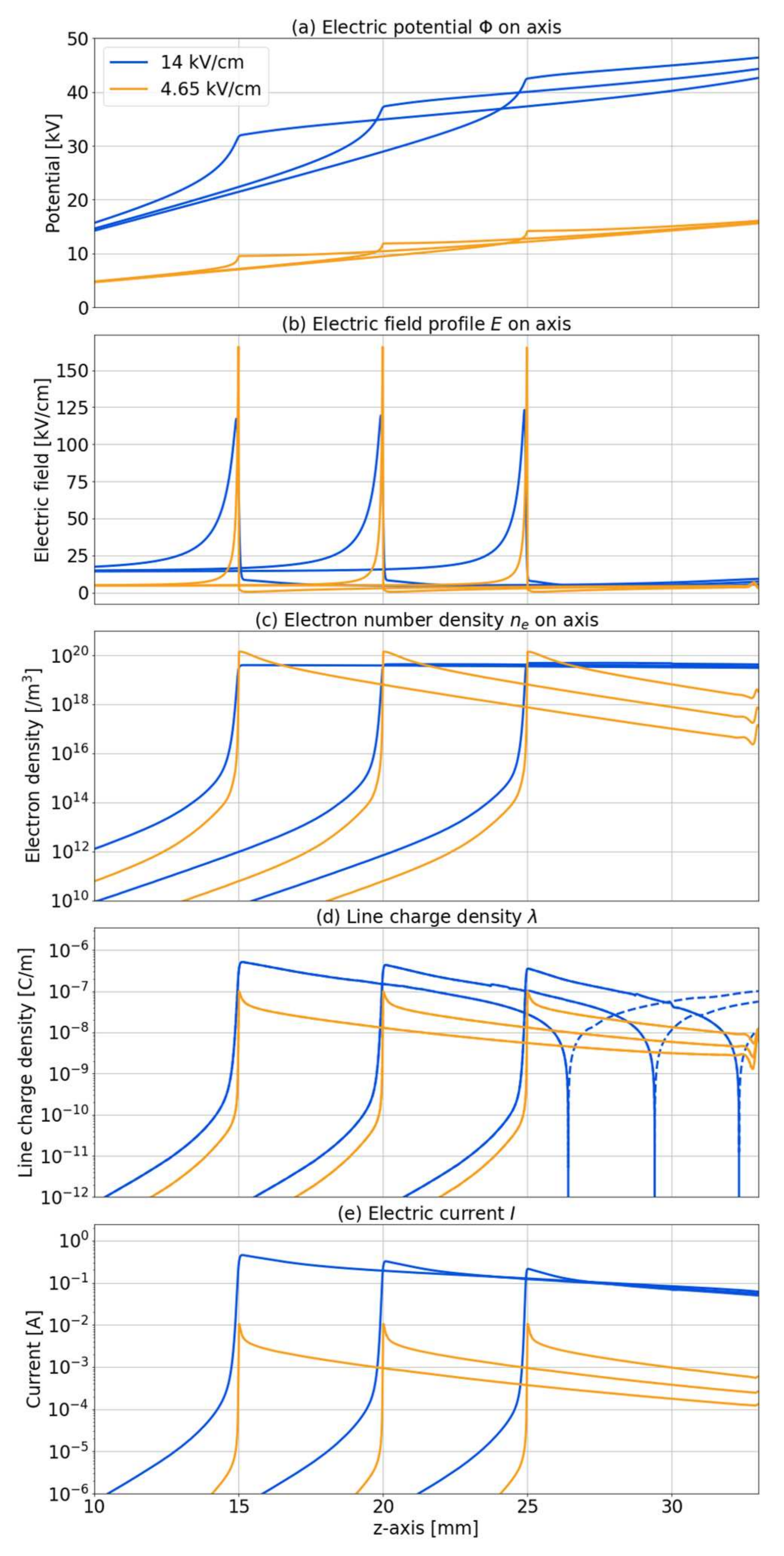}
    \caption{Axial profiles, line charge density, and current of the streamers in the same background fields of $14$ and $4.65$~kV/cm as in figure~\ref{fig:electron-density}. Here they are shown  when their maximal electric field is at $z = 15$, $20$, and $25$~mm. The panels show, from top to bottom, as a function of the axis coordinate $z$:
    (a) the electric potential $\phi$ on axis, (b) the electric field profile $E$ on axis, (c) the electron number density $n_e$ on axis, (d) the line charge density $\lambda$, which is the charge density integrated over the radial cross section (where the dashed lines represent negative values), (e) the electric current $I$, which is the current density also integrated over the radial cross section. The legend on the first panel applies to all panels.} 
    \label{fig:axis}
\end{figure}

Figure~\ref{fig:electron-density} shows the evolution of streamers in background electric fields of $4.65$ and $14$~kV/cm. The panels show the color-coded electron density together with equipotential lines in purple. For the lower field, the streamer is shown in time steps of $100$~ns, while for the higher field, in time steps of $17.5$~ns. The same streamers are presented in figure~\ref{fig:axis} showing the electric potential, the electric field and the electron density along the streamer axis, and the line charge density and the electric current. The last two are obtained by integrating the charge density and the current density across the streamer cross section. The integration was done up to $r = 5$~mm. Several basic differences can be noted between the two streamers as they propagate through the $40$~mm gap.

{\bf The solitary streamer.} $\quad$ The streamer in the $4.65$~kV/cm field grows by about an equal length within each time step of $100$~ns. The electron density is strongly reduced about $10$~mm behind the streamer head, and the electric field returns to its background value in this region and further behind, as can be seen from the straight and equidistant equipotential lines. Overall, the pattern of electron density and deflected equipotential lines is transported almost uniformly, without changes in shape. The streamer transports a constant amount of positive charge within its finite length, and there is no negative charge visible in the line charge density in Fig.~\ref{fig:axis}. We will call this streamer a solitary streamer or a uniformly translating streamer.

{\bf The accelerating streamer.} $\quad$ The streamer in the $14$~kV/cm field is shown in time steps of $17.5$~ns in Fig.~\ref{fig:electron-density}. It clearly accelerates, and its head radius increases. The electron density varies little along the whole channel for all time steps. There is electric current flowing in the order of $100$~mA along the whole channel, and the back part charges negatively while the front part accumulates positive charge - there is electric polarization along the whole channel. This is visible in the line charge density as well as in the field distortion along the whole body of the streamer channel. We will call this streamer an accelerating streamer.

Later in section~\ref{sec:behaviors} we will also discuss decelerating streamers and the fact that the solitary streamers exist only on the borderline between accelerating and decelerating streamers.

\begin{figure}
    \centering
    \includegraphics[width=\linewidth, keepaspectratio]{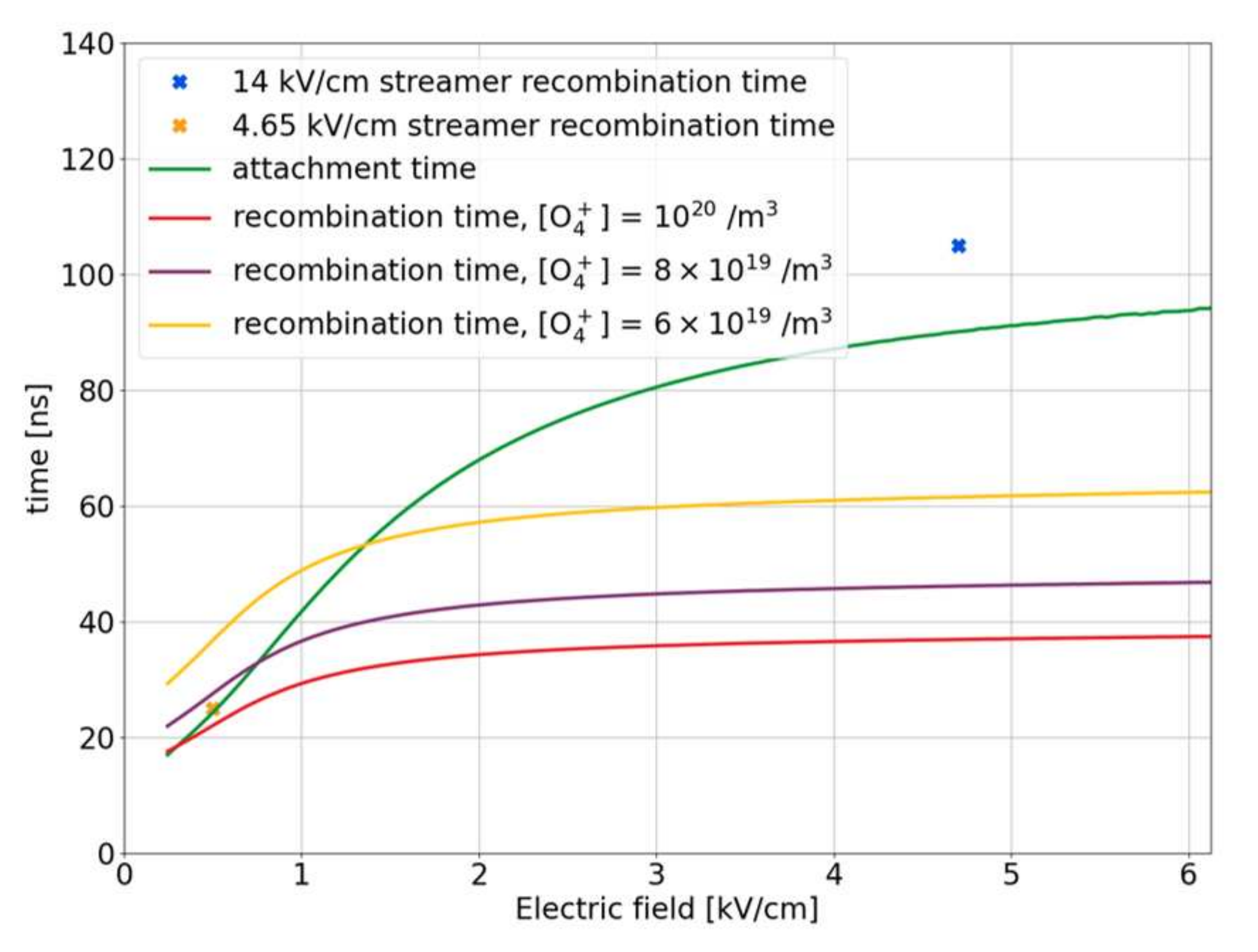}
    \caption{Attachment and electron-ion recombination time in STP air. The attachment time is plotted as a function of the electric field $E$ in green. The recombination time depends on electric field $E$ and on O$_4^+$ density, and lines for three different O$_4^+$ densities are presented. The attachment and recombination times in the channel of the $14$~kV/cm and $4.65$~kV/cm streamers are also included as blue and orange crosses for interior electric fields of $4.7$~kV/cm and $0.5$~kV/cm.}
    \label{fig:times}
\end{figure}

{\bf Attachment and recombination.} $\quad$
The lowest electric field inside the accelerating streamer is $4.7$~kV/cm, located around the middle section of the streamer channel. For the solitary streamer in the $4.65$~kV/cm background field, the electric field right behind the ionization front is as low as $0.5$~kV/cm and rises to the background value behind the solitary structure. 

The different interior electric fields and ion densities determine the attachment times - the average times until an electron attaches to an oxygen molecule, and the recombination times - the average times until an electron recombines with an O$_4^+$ ion. It should be noted here that the positive ions rapidly convert into O$_4^+$ ions. Together with the streamer velocity, these times determine over which length the streamer maintains its conductivity.

Figure~\ref{fig:times} shows some recombination times for different O$_4^+$ densities and the attachment time against the electric field. The two crosses correspond to the recombination times in the interior of the $14$~kV/cm streamer and of the $4.65$~kV/cm streamer. The recombination times and attachment times in the solitary streamer channel are as short as about $25$~ns due to the combination of low electric field and high O$_4^+$ density, while they are of the order of $105$~ns for the accelerating streamer. The high O$_4^+$ density in the solitary streamer is due to the high electric field at its tip; this high field creates a high ionization density. 

The slow propagation of the solitary streamer also gives electrons sufficient time to get attached to oxygen molecules and recombine with O$_4^+$ molecules. The accelerating streamer propagates much faster, with a higher internal field, leaving no time for electron attachment or recombination. We see in the third panel of Figure~\ref{fig:axis} that the electron density of the solitary streamer decays behind the ionization front by several orders of magnitude while the electron density in the channel of the accelerating streamer is essentially constant.

\begin{figure*}
    \centering
    \includegraphics[width=\linewidth]{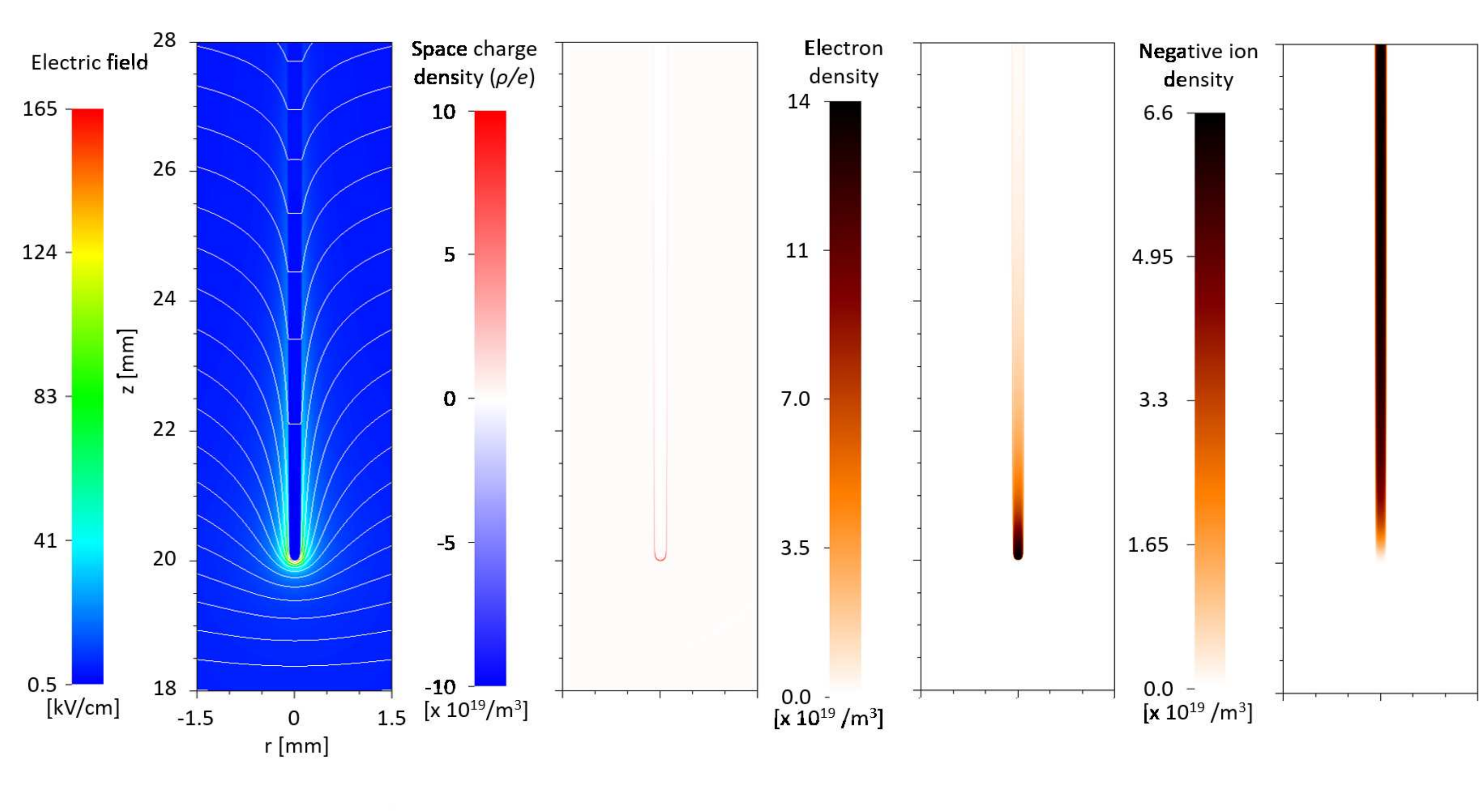}
    \caption{Plots of the $4.65$~kV/cm streamer zoomed into the streamer head when it is at $z = 20$ mm. From left to right: (1) electric field with white equipotential lines, (2) space charge density, (3) electron density, and (4) negative ion density.}
    \label{fig:head}
\end{figure*}

In Figure~\ref{fig:head}, several plots zooming in on the head of the solitary streamer are presented. The electric field inside the channel of the solitary streamer is screened to a low value, represented by the widely separated horizontal equipotential lines. Almost all the net charge is located on the streamer surface - the positive space charge layer shown in the second panel of Fig.~\ref{fig:head}. The low electric field in the streamer interior leads to fast electron attachment as discussed above, and this is evident in the electron density contour plot, where the electron density reduces in magnitude behind the streamer head. Electron attachment produces negative ions, and since recombination time and attachment time are nearly equal behind the ionization front, about half of the electrons are lost due to attachment and the other half to electron-ion recombination. Thus, the density of negative ions at the back end of the channel is about half of the electron density at the streamer head.

\subsection{Propagation modes as a function of the field \label{sec:behaviors}}

\begin{figure*}
    \centering
    \includegraphics[width=\linewidth]{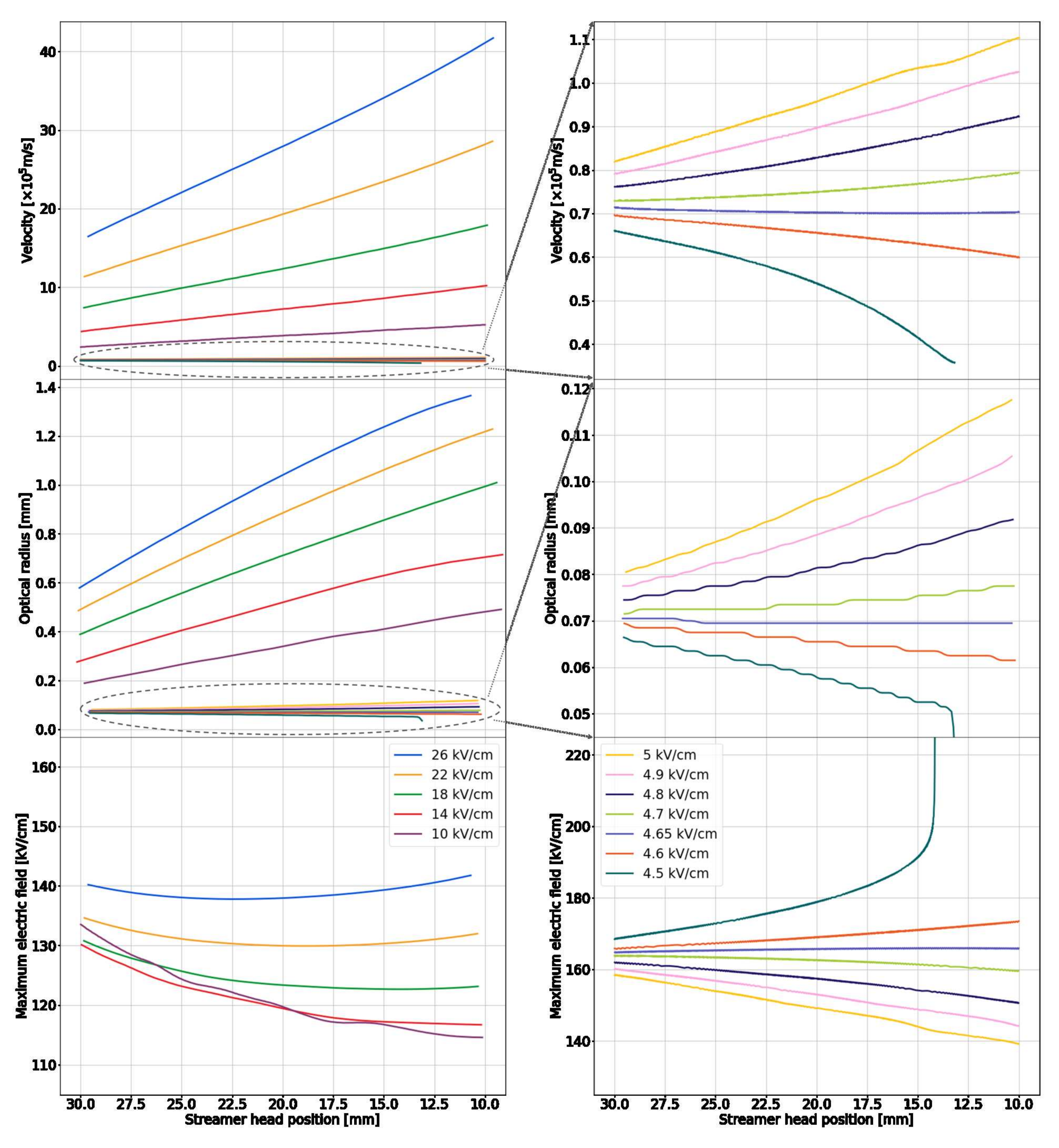}
    \caption{Properties of positive streamers as a function of length in different background fields, as indicated in the panels. Top panels show the streamer velocity, middle panels the optical radius, and bottom panels the maximum electric field. Plots on the right have a different range of values in the vertical axis and focus on the streamers in fields of $5$~kV/cm and lower. Radii have uncertainties of $\pm 1.2~\mu$m due to the finite size of the numerical grid.}
    \label{fig:properties}
\end{figure*}

Three parameter regimes can be identified in Figure~\ref{fig:properties}, which has the velocity, optical radius, and maximum electric field of the streamers as a function of length. First, there are the accelerating streamers that speed up as they lengthen, and their radius increases as they accelerate. This is the case for streamers in background electric fields above $4.65$~kV/cm. This is also the case most frequently reported and commonly observed in streamer simulations.

Second, there are uniformly propagating streamers, in a background field of $4.65$~kV/cm. They exist as a limit between accelerating and decelerating streamers, and they maintain a nearly uniform velocity. Other streamer properties such as the head radius and enhanced electric field do not change in time either. For the streamer in our simulation, the radius remained at $65~\mu$m while it was uniformly propagating. These solitary streamers can maintain their shape because they have a finite and constant length where the electron density is relevant and the electric field is modified. They carry a fixed amount of positive charge over a finite length, and therefore act as a point charge from a sufficiently far distance. The streamer is able to propagate indefinitely in this background field. This behavior can be related to the old concept of the streamer stability field, which we discuss further in section~\ref{sec:stability_field}.

Third and last, there are the decelerating streamers. We find them in fields below  $4.65$~kV/cm. Streamers in such fields slow down as they lengthen, and their head radius decreases in time while the maximum electric field increases. This happens because the electric screening of the streamer interior improves when the ionization front slows down. The decreasing radii of the decelerating streamers can be explained by the decreasing potential in the streamer head due to voltage lost in the streamer channel~\cite{starikovskiy2021}. Some of our simulated decelerating streamers do not manage to cross the domain, as shown by the case of the streamer with a background field of $4.5$~kV/cm. The streamer decelerated and eventually stagnated with a streamer radius of $49~\mu$m. This stagnating behavior was described earlier in~\cite{Pancheshnyi_2004, starikovskiy2021, Starikovskiy_2020} and observed experimentally in~\cite{briels_positive_2008, Qin_2014}. Numerically, we observe that the simulation time steps, which are usually in picoseconds, drop by two orders of magnitude because the maximum electric field values suddenly increase to magnitudes greater than $300$~kV/cm in a very small region ahead of the ionization front. One reason for this numerical instability may be the artificial diffusion of electrons from the channel to the high-field region ahead of the streamer tip~\cite{Teunissen_2020}. The physical process of streamer stagnation was always accompanied by such numerical instabilities in our simulations.

Although we used different initial conditions depending on the applied electric field, we still expect actual streamers in low background fields to grow in a similar manner as we've identified. Streamers are characterized by their velocity, radius, and maximal electric field, which determine how they propagate. As long as they share the same properties as our results, their dynamics would be the same. Additionally, in \cite{Pancheshnyi_2004} it was observed that beyond $1$~cm from the point of initiation, the initial condition is forgotten by the streamer.

\subsection{Nonlinear dependence of field enhancement and plasma chemistry on the background field}

\begin{figure}
    \centering
    \includegraphics[width=\linewidth]{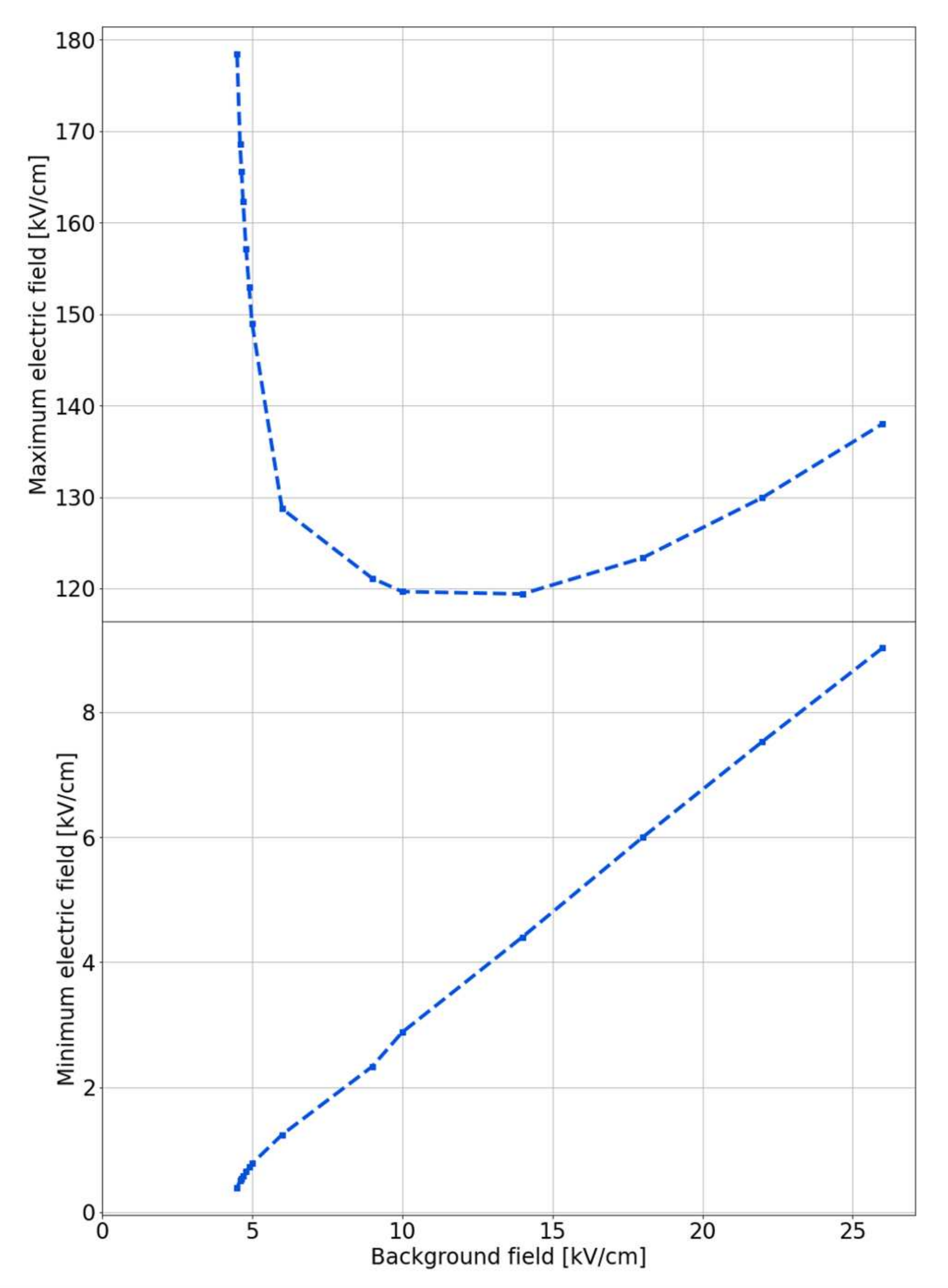}
    \caption{Maximal (top) and minimal (bottom) electric field in the streamer as a function of the background electric field. The maximal field is measured at the streamer head while the minimal field is from behind the streamer ionization front. The values were acquired when the streamer heads were at position $z = 20$~mm.}
    \label{fig:fields}
\end{figure}

The streamer dynamics nonlinearly depend on the background electric field $E_\textrm{back}$. In the top panel of Figure~\ref{fig:fields}, we see the maximal field $E_\textrm{max}$ as a function of the background field $E_\textrm{back}$, evaluated at the moment when the streamer heads are at $z=20$~mm. The curve has a minimum of about $E_\textrm{max}=120$~kV/cm for a background electric field around $E_\textrm{back}=12$~kV/cm. For $E_\textrm{back}$ increasing up to $26$~kV/cm, the maximal field increases up to $140$~kV/cm, while below $10$~kV/cm the maximal electric field increases rapidly, until it diverges for $E_\textrm{back}=4.5$~kV/cm. As the electron energy distribution and the induced plasma chemistry depend on the electric field configuration, we conclude that the plasma chemistry could also depend nonlinearly on the background electric field. This observation requires further investigation in the future. 

The minimum electric field behind the ionization front of the streamers as a function of the background field is presented in the bottom panel of Figure~\ref{fig:fields}. We found that the minimum electric field inside the streamer channel depends almost linearly on the background electric field. It vanishes for the stagnating streamer, and it reaches $9$~kV/cm for $E_\textrm{back}=26$~kV/cm.

\subsection{Heating}

In \cite{Komuro_2014} a streamer that propagated for a few hundred nanoseconds was found to already heat the gas significantly. As the solitary streamer also took a couple hundred nanoseconds to cross the computational domain, we evaluated the temperature increase. We used the expression $Q = \int
\textbf{j} \cdot \textbf{E} \: dt$~\cite{Agnihotri_2017} to calculate the deposited electric energy density $Q$;
here $\textbf{j}$ is the electric current density. 
Even if we assume that the full deposited energy is converted into heat, the temperature on the axis of the solitary streamer increases only by 6~K after 400~ns. 

The difference with the result of~\cite{Komuro_2014} lies in the fact that the energy deposition per electron is not determined by time, but by the distance the electron travels in the electric field. In the solitary streamer the electrons attach or recombine after a short propagation distance. It should be noted though that the electron density is higher in the head of the solitary streamer than in a higher background electric field.

\subsection{Ion motion \label{sec:ion_motion}}

As electrons attach to oxygen and form negative ions in the channel, we briefly explore the effect of ion motion on streamer behavior. Incorporating ion motion in streamer simulations with $14$~kV/cm and $9$~kV/cm background fields did not visibly change anything in the results. For these cases, the streamer still propagates fast enough that ion motion has negligible effects. We only start to observe effects in low background electric fields, when enough time is available to deplete the electron density through attachment and recombination.

\begin{figure}
    \centering
    \includegraphics[width=\linewidth]{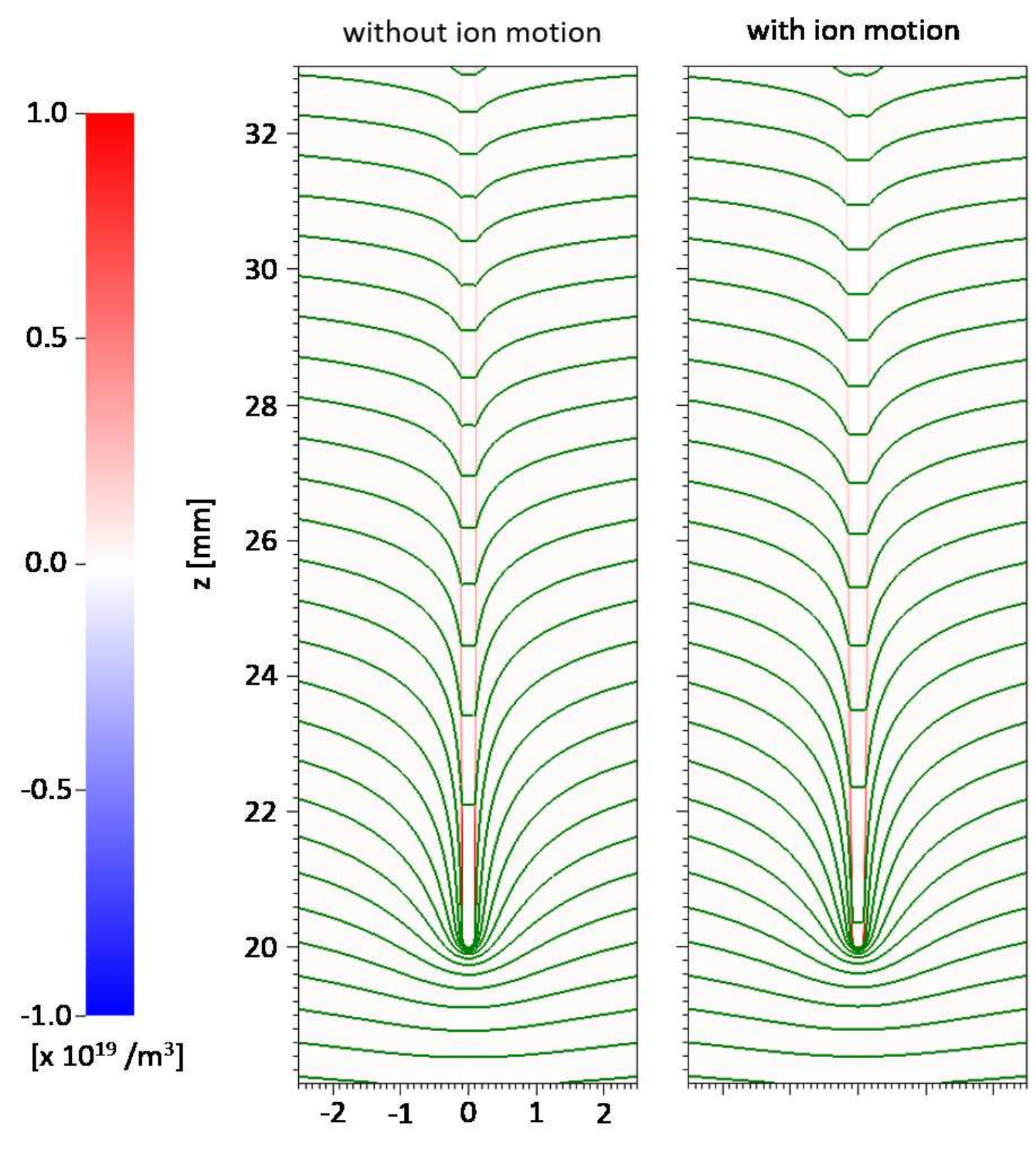}
    \caption{Charge density of the $4.65$~kV/cm streamer without ion motion (left) and with ion motion (right), including green equipotential lines.}
    \label{fig:ion_contour}
\end{figure}

Figure~\ref{fig:ion_contour} shows the total charge density of the solitary streamer with and without ion motion. We see that the channel of the streamer with ion motion is wider at the back. The space charge layer of these streamers is made up of positive ions, and without ion motion they remain fixed in space. Only reactions can change the densities of these ions in time. With the inclusion of ion motion, these ions are now moving radially outward in response to the electric field they are subjected to. The ion drift in the local field also causes the streamer head to lose some focus, leading to slower propagation. We observe similar phenomena in negative streamers, whose space charge layers are made up of the very mobile electrons.

\begin{figure}
    \centering
    \includegraphics[width=\linewidth]{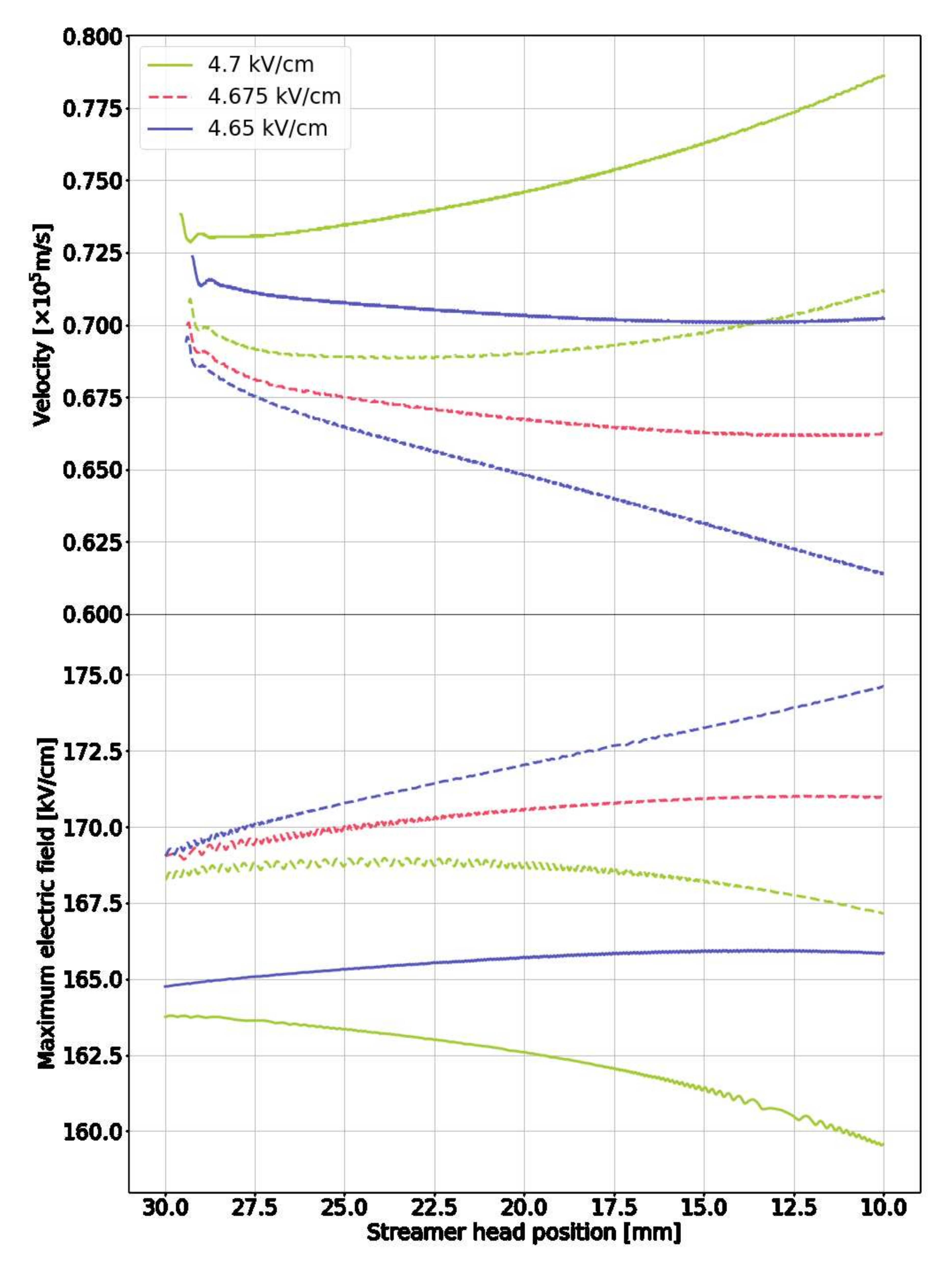}
    \caption{Velocity (top panel) and maximum electric field (bottom panel) of streamers against streamer head position for simulations with and without ion motion. Broken lines stand for simulations with ion motion. The legend on the top panel applies to the bottom panel as well.}
    \label{fig:ion_motion}
\end{figure}

When ion motion is included, the streamer propagates more slowly. This can be observed in the upper panel of Figure~\ref{fig:ion_motion}, where streamers in the same background field with and without ion motion are presented. The previously discovered uniformly translating streamer at $4.65$~kV/cm decelerates when ion motion is included in the simulation. A new background electric field for uniform translation was found at $4.675$~kV/cm, only slightly higher than the previous background field, with a slightly lower uniform velocity of $0.66 \times 10^{5}$~m/s. Thus, the effect of ion motion on streamer dynamics does not appear to be strong in this case. We will be using this new uniformly translating streamer for our comparisons in section~\ref{sec:validation}.

The maximal electric field of the streamers in fields of $4.65$ to $4.7$~kV/cm with and without ion motion is plotted as a function of the streamer head position in the lower panel of Figure~\ref{fig:ion_motion}. When ion motion is included, the maximal electric fields at the same background electric field are higher, which is consistent with the smaller head radii. 

Finally, we compare the maximal electron drift velocity with the velocity of the uniformly translating streamer with ion motion. The maximal electric field at the streamer head has a constant value of $171$~kV/cm, which gives us an electron drift velocity of $5.3 \times 10^{5}$~m/s, while the streamer velocity is $6.6\times 10^4$~m/s - almost an order of magnitude smaller than the drift velocity. This is possible for positive streamers, where these velocities are directed in opposite directions, but not for negative streamers.

\section{Comparison with experiments} \label{sec:validation}

\subsection{The stability field \label{sec:stability_field}}

Recently, the concept of the streamer stability field has been more commonly used in association with streamers propagating in inhomogeneous electric fields. It relates the maximum length a streamer could gain to the applied voltage~\cite{Allen_1995, allen1991, Babaeva_1996, seeger_2018, Veldhuizen_2002}. An older definition used the term stability field to mean the homogeneous electric field in which a streamer would propagate in a stable manner - without changes in velocity and shape~\cite{phelps_fieldenchanced, griffiths_effectofairpressure, gallimberti_longspark}.

If we only consider the streamer channel length as the length behind the streamer head with substantial electron density, we observe the solitary streamer to have a constant length as it propagates. The solitary streamer has a uniform shape, and it follows, using the older definition of the stability field, that the solitary streamer is propagating in the stability field of STP dry air at $4.675$~kV/cm. This value agrees with the measured stability field of $4$~kV/cm in experiments~\cite{griffiths_effectofairpressure, gallimberti_longspark} for the original definition.
With the newer definition, the stability field is reported to be between 4.5-5~kV/cm~\cite{Allen_1995, razer1991}.

\subsection{Radius and velocity of solitary and minimal streamers.}

In the pin-to-plate experiments of \cite{Briels_2006}, it was found that after several branching events or in a quite weak field, streamers would approach a minimal diameter, and they were called minimal streamers. The solitary streamers are essentially the thinnest streamers that we found in our simulations as the stagnating streamers are not much thinner and hardly emit any light. Therefore we now compare their properties.

The simulated solitary streamer that includes ion motion has a radius of $55~\mu$m, and this value is not far from the experimental findings in \cite{nijdam_probing_2010}, which give $65~\mu$m as the minimal streamer diameter in $1$~bar air. The uniform velocity of our solitary streamer is $0.7 \times 10^{5}$~m/s, which falls in the range of the measured velocity of $\left(0.5 - 1\right) \times 10^5$~m/s of minimal streamers. Therefore we can conclude that the simulations match the experiments within 20\%.

\section{Conclusions and Outlook \label{sec:conclusion}}

We simulated single positive streamers in air at standard temperature and pressure in homogeneous background fields ranging from $4.5$~kV/cm to $26$~kV/cm in a $4$~cm gap, and we came to the following conclusions:
\begin{itemize}
    \item[1.] The solitary streamer (or uniformly translating streamer) with dominant electron attachment and recombination behind the head lays a theoretical basis for the much used concept of a stability field. Streamers in higher fields increase in radius and velocity, while the solitary streamer transports a fixed amount of positive charge that is substantial only over a finite length.
    \item[2.] The solitary streamer motion explains how a streamer can propagate over distances in meter length-scales though the conductivity of the back part of the channel disappearing due to attachment and recombination. The velocity of such a streamer can be an order of magnitude smaller than the electron drift velocity in its maximal electric field.
    \item[3.] The value of the stability field of $4.675$~kV/cm in our simulations in STP air agrees well experimentally measured values.
    \item[4.] Minimal streamers are the thinnest and slowest streamers that have been experimentally observed~\cite{Briels_2006}. Our values for the optical radius and velocity of solitary streamers agree well with measurements of these so-called minimal streamers. Even better agreement could possibly be found if for example humidity, repetition rate, and fluid model limitations were taken into account.
    \item[5.] The solitary streamer causes negligible gas heating even after propagating for several hundreds of nanoseconds.
    \item[6.] Ion motion plays a minor role for solitary streamers, but its effect increases as streamers slow down.
    \item[7.] The maximal electric field at the streamer head is not a monotonic function of the background field, but it has a minimum for a background field of about $12$~kV/cm. The implications of this on the electron energy distribution and on the optimization of the plasma chemistry will need to be investigated.
\end{itemize}

Future research could look into model reduction based on the solitary streamer, as it does not depend on time in a co-moving frame. How our current findings translate to other gases with different plasma-chemical reactions and photoionization rates also merits further investigation. There is an avenue for exploring the behavior of accelerating streamers on longer timescales, and the existence of the solitary positive streamer also raises the question of whether the solitary mode of propagation could also be observed in negative streamers.
Finally, another open question is how and when solitary streamers form in background fields with a spatial gradient, as is common in experiments.

\section{Acknowledgments}
We thank Andy Martinez and Xiaoran Li for the help with post-processing methods. H.F. was funded by the European Union's Horizon 2020 research and innovation programme under the Marie Skłodowska-Curie grant agreement SAINT 72233.

\section*{References}
\bibliographystyle{iopart-num}
\bibliography{refs}

\providecommand{\noopsort}[1]{}\providecommand{\singleletter}[1]{#1}%
\providecommand{\newblock}{}
\begin{thebibliography}{10}
\expandafter\ifx\csname url\endcsname\relax
  \def\url#1{{\tt #1}}\fi
\expandafter\ifx\csname urlprefix\endcsname\relax\def\urlprefix{URL }\fi
\providecommand{\eprint}[2][]{\url{#2}}

\bibitem{Babaeva1997}
Babaeva N and Naidis G 1997 {\em IEEE Transactions on Plasma Science\/} {\bf
  25} 375--379

\bibitem{Kulikovsky1998}
Kulikovsky A~A 1998 {\em Phys. Rev. E\/} {\bf 57}(6) 7066--7074
  \urlprefix\url{https://link.aps.org/doi/10.1103/PhysRevE.57.7066}

\bibitem{Pancheshnyi_2005}
Pancheshnyi S, Nudnova M and Starikovskii A 2005 {\em Phys. Rev. E\/} {\bf
  71}(1) 016407
  \urlprefix\url{https://link.aps.org/doi/10.1103/PhysRevE.71.016407}

\bibitem{ebert_2010}
Ebert U, Nijdam S, Li C, Luque A, Briels T and van Veldhuizen E 2010 {\em
  Journal of Geophysical Research: Space Physics\/} {\bf 115} A00E43

\bibitem{Nijdam_review_2020}
Nijdam S, Teunissen J and Ebert U 2020 {\em Plasma Sources Sci. Technol.\/}
  {\bf 29} 103001 \urlprefix\url{https://doi.org/10.1088/1361-6595/abaa05}

\bibitem{laroussi2014}
Laroussi M 2014 {\em Plasma Processes and Polymers\/} {\bf 11} 1138--1141
  \urlprefix\url{https://doi.org/10.1002/ppap.201400152}

\bibitem{starikovskaia_plasma-assisted_2014}
Starikovskaia S~M 2014 {\em J. Phys. D: Appl. Phys.\/} {\bf 47} 353001 ISSN
  1361-6463 \urlprefix\url{http://dx.doi.org/10.1088/0022-3727/47/35/353001}

\bibitem{bardos2010}
B{\'a}rdos L and Bar{\'a}nkov{\'a} H 2010 {\em Thin Solid Films\/} {\bf 518}
  6705--6713 \urlprefix\url{https://doi.org/10.1016/j.tsf.2010.07.044}

\bibitem{Allen_1995}
Allen N~L and Ghaffar A 1995 {\em Journal of Physics D: Applied Physics\/} {\bf
  28} 331--337 \urlprefix\url{https://doi.org/10.1088/0022-3727/28/2/016}

\bibitem{phelps_fieldenchanced}
Phelps C~T 1971 {\em Journal of Geophysical Research (1896-1977)\/} {\bf 76}
  5799--5806 \urlprefix\url{https://dx.doi.org/10.1029/JC076i024p05799}

\bibitem{briels_positive_2008}
Briels T~M~P, Kos J, Winands G~J~J, van Veldhuizen E~M and Ebert U 2008 {\em J.
  Phys. D: Appl. Phys.\/} {\bf 41} 234004 ISSN 1361-6463
  \urlprefix\url{http://dx.doi.org/10.1088/0022-3727/41/23/234004}

\bibitem{Luque_branching_2011}
Luque A and Ebert U 2011 {\em Phys. Rev. E\/} {\bf 84}(4) 046411
  \urlprefix\url{https://link.aps.org/doi/10.1103/PhysRevE.84.046411}

\bibitem{Pancheshnyi_2004}
Pancheshnyi S~V and Starikovskii A~Y 2004 {\em Plasma Sources Science and
  Technology\/} {\bf 13} B1--B5
  \urlprefix\url{https://doi.org/10.1088/0963-0252/13/3/b01}

\bibitem{starikovskiy2021}
Starikovskiy A~Y, Aleksandrov N~L and Shneider M~N 2021 {\em Journal of Applied
  Physics\/} {\bf 129} 063301 \urlprefix\url{https://doi.org/10.1063/5.0037669}

\bibitem{francisco2021}
Francisco H, Bagheri B and Ebert U 2021 {\em Plasma Sources Science and
  Technology\/} {\bf 30} 025006
  \urlprefix\url{https://doi.org/10.1088/1361-6595/abdaa3}

\bibitem{griffiths_effectofairpressure}
Griffiths R~F and Phelps C~T 1976 {\em Quarterly Journal of the Royal
  Meteorological Society\/} {\bf 102} 419--426
  \urlprefix\url{https://dx.doi.org/10.1002/qj.49710243211}

\bibitem{gallimberti_longspark}
{Gallimberti, I} 1979 {\em J. Phys. Colloques\/} {\bf 40} C7--193--C7--250
  \urlprefix\url{https://doi.org/10.1051/jphyscol:19797440}

\bibitem{hagelaar_solving_2005}
Hagelaar G~J~M and Pitchford L~C 2005 {\em Plasma Sources Science and
  Technology\/} {\bf 14} 722--733 ISSN 1361-6595
  \urlprefix\url{http://dx.doi.org/10.1088/0963-0252/14/4/011}

\bibitem{Pancheshnyi_effective_2013}
Pancheshnyi S 2013 {\em Journal of Physics D: Applied Physics\/} {\bf 46}
  155201 \urlprefix\url{https://doi.org/10.1088/0022-3727/46/15/155201}

\bibitem{Aleksandrov_ionization_1999}
Aleksandrov N~L and Bazelyan E~M 1999 {\em Plasma Sources Science and
  Technology\/} {\bf 8} 285--294
  \urlprefix\url{https://doi.org/10.1088\%2F0963-0252\%2F8\%2F2\%2F309}

\bibitem{kossyi_kinetic_1992}
Kossyi I~A, Kostinsky A~Y, Matveyev A~A and Silakov V~P 1992 {\em Plasma
  Sources Sci. Technol.\/} {\bf 1} 207--220 ISSN 1361-6595
  \urlprefix\url{http://dx.doi.org/10.1088/0963-0252/1/3/011}

\bibitem{luque2017streamer}
Luque A, Gonz{\'a}lez M and Gordillo-V{\'a}zquez F 2017 {\em Plasma Sources
  Science and Technology\/} {\bf 26} 125006

\bibitem{phelps_anisotropic_1985}
Phelps A~V and Pitchford L~C 1985 {\em Phys. Rev. A\/} {\bf 31} 2932--2949
  \urlprefix\url{http://dx.doi.org/10.1103/PhysRevA.31.2932}

\bibitem{phelps_data}
Phelps database, retrieved March 2019 \urlprefix\url{www.lxcat.net}

\bibitem{zheleznyak_photoionization_1982}
Zheleznyak M~B, Mnatsakanian A~K and Sizykh S~V 1982 {\em Teplofizika Vysokikh
  Temperatur\/} {\bf 20} 423--428

\bibitem{bourdon_efficient_2007}
Bourdon A, Pasko V~P, Liu N~Y, Célestin S, Ségur P and Marode E 2007 {\em
  Plasma Sources Sci. Technol.\/} {\bf 16} 656--678 ISSN 1361-6595
  \urlprefix\url{http://dx.doi.org/10.1088/0963-0252/16/3/026}

\bibitem{luque_photoionization_2007}
Luque A, Ebert U, Montijn C and Hundsdorfer W 2007 {\em Appl. Phys. Lett.\/}
  {\bf 90} 081501 ISSN 0003-6951
  \urlprefix\url{http://dx.doi.org/10.1063/1.2435934}

\bibitem{Tochikubo_2002}
Tochikubo F and Arai H 2002 {\em Japanese Journal of Applied Physics\/} {\bf
  41} 844--852 \urlprefix\url{https://doi.org/10.1143\%2Fjjap.41.844}

\bibitem{teunissen_simulating_2017}
Teunissen J and Ebert U 2017 {\em Journal of Physics D: Applied Physics\/} {\bf
  50} 474001 ISSN 1361-6463
  \urlprefix\url{http://dx.doi.org/10.1088/1361-6463/aa8faf}

\bibitem{Teunissen_afivo_2018}
Teunissen J and Ebert U 2018 {\em Computer Physics Communications\/} {\bf 233}
  156–166 ISSN 0010-4655
  \urlprefix\url{http://dx.doi.org/10.1016/j.cpc.2018.06.018}

\bibitem{luque_density_2012}
Luque A and Ebert U 2012 {\em Journal of Computational Physics\/} {\bf 231}
  904--918 ISSN 0021-9991
  \urlprefix\url{http://dx.doi.org/10.1016/j.jcp.2011.04.019}

\bibitem{Pancheshnyi_2000}
Pancheshnyi S, Starikovskaia S and Starikovskii A 2000 {\em Chemical Physics\/}
  {\bf 262} 349--357 ISSN 0301-0104

\bibitem{bagheriSimulationPositiveStreamers2020}
Bagheri B, Teunissen J and Ebert U 2020 {\em Plasma Sources Science and
  Technology\/} {\bf 29} 125021
  \urlprefix\url{https://doi.org/10.1088/1361-6595/abc93e}

\bibitem{Starikovskiy_2020}
Starikovskiy A~Y and Aleksandrov N~L 2020 {\em Plasma Sources Science and
  Technology\/} {\bf 29} 075004
  \urlprefix\url{https://doi.org/10.1088/1361-6595/ab9484}

\bibitem{Qin_2014}
Qin J and Pasko V~P 2014 {\em Journal of Physics D: Applied Physics\/} {\bf 47}
  435202 \urlprefix\url{https://doi.org/10.1088/0022-3727/47/43/435202}

\bibitem{Teunissen_2020}
Teunissen J 2020 {\em Plasma Sources Science and Technology\/} {\bf 29} 015010
  ISSN 1361-6595 \urlprefix\url{http://dx.doi.org/10.1088/1361-6595/ab6757}

\bibitem{Komuro_2014}
Komuro A and Ono R 2014 {\em Journal of Physics D: Applied Physics\/} {\bf 47}
  155202 \urlprefix\url{https://doi.org/10.1088/0022-3727/47/15/155202}

\bibitem{Agnihotri_2017}
Agnihotri A, Hundsdorfer W and Ebert U 2017 {\em Plasma Sources Science and
  Technology\/} {\bf 26} 095003
  \urlprefix\url{https://doi.org/10.1088/1361-6595/aa8571}

\bibitem{allen1991}
Allen N and Boutlendj M 1991 {\em Science, Measurement and Technology, IEE
  Proceedings A\/} {\bf 138} 37 -- 43

\bibitem{Babaeva_1996}
Babaeva N~Y and Naidis G~V 1996 {\em Journal of Physics D: Applied Physics\/}
  {\bf 29} 2423--2431
  \urlprefix\url{https://doi.org/10.1088/0022-3727/29/9/029}

\bibitem{seeger_2018}
Seeger M, Votteler T, Ekeberg J, Pancheshnyi S and Sánchez L 2018 {\em IEEE
  Transactions on Dielectrics and Electrical Insulation\/} {\bf 25} 2147--2156

\bibitem{Veldhuizen_2002}
van Veldhuizen E~M and Rutgers W~R 2002 {\em Journal of Physics D: Applied
  Physics\/} {\bf 35} 2169--2179
  \urlprefix\url{https://doi.org/10.1088/0022-3727/35/17/313}

\bibitem{razer1991}
Raizer Y 1991 {\em Gas Discharge Physics\/} (Springer-Verlag Berlin Heidelberg)
  ISBN 978-3-642-64760-4

\bibitem{Briels_2006}
Briels T~M~P, Kos J, van Veldhuizen E~M and Ebert U 2006 {\em Journal of
  Physics D: Applied Physics\/} {\bf 39} 5201--5210
  \urlprefix\url{https://doi.org/10.1088/0022-3727/39/24/016}

\bibitem{nijdam_probing_2010}
Nijdam S, van~de Wetering F~M~J~H, Blanc R, van Veldhuizen E~M and Ebert U 2010
  {\em J. Phys. D: Appl. Phys.\/} {\bf 43} 145204 ISSN 1361-6463
  \urlprefix\url{http://dx.doi.org/10.1088/0022-3727/43/14/145204}

\end{thebibliography}

\end{document}